\newcommand{\rmi}{\mathrm{i}}
\newcommand{\rmd}{\mathrm{d}}

\documentclass[12pt]{article}
\usepackage[dvips]{graphicx}
\usepackage{framed}
\usepackage{amsfonts}
\usepackage{array}
\usepackage{verbatim}
\usepackage{psfrag}
\usepackage{authblk}

\title{Current Density Imaging through Acoustically Encoded Magnetometry: A 
Theoretical Exploration}

\author{D. Sheltraw%
  \thanks{email: \texttt{sheltraw@berkeley.edu}}}
\affil{Henry H. Wheeler Jr. Brain Imaging Center, University of California, 
Berkeley, CA, 94720, USA.}

\date{\today}

\begin{document}
\maketitle

\begin{abstract}
The problem of determining a current density confined to a volume from 
measurements of the magnetic field it produces exterior to that volume 
is known to have non-unique solutions. To uniquely determine the current
density, or the non-silent components of it, additional spatial encoding of the 
electric current is needed. In biological systems such as the brain and heart, 
which generate electric current associated with normal function, a reliable 
means of generating such additional encoding, on a spatial and temporal scale 
meaningful to the study of such systems, would be a boon for research. 
This paper explores a speculative method by which the required additional 
encoding might be accomplished, on the time scale associated with the
propagation of sound across the volume of interest, by means of the 
application of a radially encoding pulsed acoustic spherical wave. 
\end{abstract}

\maketitle

\section{Introduction}
 
The magnetic inverse problem consists of the estimation of the magnetic source, an electric current density, given the field it produces external to the 
source. However, as was shown long ago by Helmholtz \cite{HvH}, a current
distribution inside a conductor cannot in general be determined uniquely 
from knowledge of the electromagnetic field exterior to the conductor. 
It is well known that some current configurations will be "silent" in that they
produce no electric or magnetic field on the surface of a sphere (the 
measurement shell) containing the current density. But this is not the only 
reason for the lack of invertibility. It is also known that even the non-silent 
currents can not be uniquely obtained through an inversion of the 
electromagnetic field data \cite{FKM,SC}. To obtain a unique solution of the
electromagnetic inverse problem additional spatial encoding beyond that 
obtained from electromagnetic measurements taken outside the brain or heart 
is needed. 

The measurement of current density within a mechanically inaccessible 
volume is an important topic in biology. Magnetoencephalography (MEG), 
electroencephalography (EEG), magnetocardiography (MCG), and 
electrocardiography (ECG) have for decades been used as a means to study brain 
and heart function. In the associated literature the additional spatial 
encoding needed to obtain a unique inversion is supplied by means of 
mathematical or physical constraints for which there exists no independent 
justification or no reasonable estimates of error bounds. The result is a 
current density image of unknown veracity at best.  

Instead of trying to supplement the electromagnetic field measurements with 
additional mathematical or physical constraints some researchers have 
attempted to use other means to encode the missing spatial information. 
Haider {\it et al.} \cite{HHX} has presented a method by which measurement of 
the electric field external to a time-varying current density in combination 
with a spatially selective acoustic wave might be used to create a current 
density image. Witte {\it et al.} \cite{Witte_2,Witte_1}, again working with
measurements of the electric field, have investigated using the 
acousto-electric effect, wherein an applied acoustic wave produces a change
in the resistivity of a medium, as a means of imaging current density. 
However, their work appears to neglect the effect of the acoustic wave
on the primary current and does not make clear what components or properties
of the current density can be uniquely determined.

The purpose of this paper is to present a theoretical exploration of a novel 
method which could potentially provide a unique inversion of the non-silent 
current density source from magnetic field data measured on a spherical 
shell in the presence of a spatially selective acoustic spherical wave.
In particular an emphasis will be placed upon the imaging of current density 
within the brain (generated by neuronal activity) or a brain-like model system. 
This paper will not be concerned with how the acoustic wave might be generated 
and applied but various means for doing so, from standard ultrasound 
transducers to photoacoustic \cite{PA_rev} interaction with biological tissue, 
exist. Although implementation of this method in the neuroscience setting 
would be challenging because of sensitivity and noise issues, and possibly 
because of issues with the proposed current density model, this work 
still has value in that it helps bring into focus some of the theoretical 
issues associated with current density imaging. 

It has been shown \cite{FKM,SC} that knowledge of the magnetic field on the 
surface of a spherical shell enclosing a compactly supported current density
is specifically lacking (apart from the silence of some current) with 
respect the radial encoding of the current density. To obtain a radial encoding
of the non-silent current density the
method explored in this paper uses a spherical acoustic wave. In brief, as 
the acoustic wave propagates radially through the medium supporting the current 
density it perturbs the current density, and its magnetic field, in manner 
that depends upon the shape of the acoustic wave packet and its position (The 
speed of the wave is assumed to be known). When the magnetic field is sampled
as the acoustic wave propagates through the medium sufficient radial encoding
information may be obtained to yield a unique (in principal) inversion of the
non-silent current density. The time scale associated with the acquisition of 
the data used to generate an image of the current density would be roughly 
equal to the time for the propagation of an acoustic wave across the volume of 
interest plus some additional time to allow reflecting waves to dissipate. For 
a system of brain-like spatial extent (speed of sound in brain is approximately 
1500 m/s) this would mean a time scale significantly less than a millisecond 
although signal averaging would most certainly be needed to suppress noise.

This paper makes use of a highly idealized model of the brain which hopefully
incorporates enough realism to spark interest in further development of the 
proposed method. The model consists of a set of assumptions about the 
geometry, mechanical properties and electromagnetic properties of the system. 
These assumptions include but are not limited to: (1) The medium in which the 
current density is supported has uniform isotropic homogeneous acoustic 
properties, (2) The mass density of the medium is uniform on the compact 
support of the current density, (3) The conductivity of the medium is 
spherically symmetric, (4) The attenuation and dispersion of the applied 
acoustic wave may be neglected, (5) The current density does not vary in time 
and (6) The total charge density on the macroscopic scale is zero at all times.
Other assumptions will be introduced in a timely manner as the interaction of 
the acoustic wave with the medium is explored. The assumptions associated with 
this simple model are obvious shortcomings of this work but future efforts 
may be able to produce a more realistic model without sacrificing the essential
features of the proposed method.

The paper is organized as follows. In Section \ref{us_wave} a model is
introduced which is hoped to adequately describe the interaction of an 
acoustic spherical wave with a brain-like medium that supports an electric 
current density source. In Section \ref{mag_field} an expression is derived 
(similar to results given in \cite{SC, FGK}) which makes clear that a limited 
set of radial moments associated with the current density can be determined 
from the magnetic field data alone. So that the reader may better appreciate
the present state of MEG "imaging" it is also shown that most of the data 
needed to uniquely determine a current density will, for spatial resolutions 
of interest, be fixed by additional encoding or constraining data. In Section 
\ref{nonsilent} it is shown how magnetic field data obtained during application
of an acoustic pulsed spherical wave could be used to uniquely obtain the 
current density of this model. In Section \ref{sec-sigmag} signal strength and 
noise considerations are addressed. Finally Section \ref{sec-disc} concludes 
with a brief discussion of some of the obstacles associated with the proposed 
method. Topics for further exploration are suggested throughout the paper.
 
\section{Acoustic Wave in a Model Conductor}
\label{us_wave}

The current density within a biological-like medium of tensor conductivity 
$\sigma_{ij}({\bf r})$ may be characterized as the sum of a {\it primary 
current} ${\bf J}_p({\bf r})$ and a {\it volume current} ${\bf J}_v({\bf r}) 
= \sum_{ij} {\hat{\bf x}}_i \sigma_{ij}({\bf r}) E_j({\bf r})$ (magnetization 
current density and polarization current density can usually be neglected)
where $E_j({\bf r})$ are the Cartesian components of an electric field. This 
distinction between primary and secondary currents is associated with a choice 
of spatial scale. A conductivity is always associated with a particular scale. 
In this case, where a distinction is made between primary and volume current, 
the conductivity will be associated with a macroscopic scale or volume $V_m$ 
that might contain on the order of thousands of neurons (The range of pyramidal
cell density in human cortex is $5,000 -80,000$ mm\textsuperscript{-3} and 
depends on cortical region \cite{Econo}.). Since the conductivity is defined 
on this macroscopic scale then the Ohm's Law current must be given by its 
product with an electric field of equal spatial scale. There will however be 
currents on the microscopic scale (perhaps a scale of a few neurons) that may 
be given by Ohm's Law as well. However the Ohm's Law expression on the 
microscopic scale will only be correct when it is expressed in terms of the 
electric field on the microscopic scale. The primary current can therefore be 
seen as a macroscopic quantity that is given by averaging over $V_m$ the 
product of the microscopic conductivity and microscopic electric field.

When the macroscopic conductivity is spherically symmetric, $\sigma_{ij}
({\bf r}) = \sigma(r)\delta_{ij}$, it can be shown \cite{HHIL} that the radial 
component of the magnetic field exterior to a region supporting the current 
density is independent of the volume currents. We will assume that the 
conductivity is spherically symmetric prior to the application of the acoustic 
wave. Since the acoustic wave is assumed to be spherical then
the spherical symmetry of the conductivity should be maintained at all times 
and the volume currents will not contribute to the radial component of the
external magnetic field. Therefore we need only concern ourselves with the
primary current in the remainder of this paper.

The source of the signals measured in magnetoencephalography is predominantly 
the primary current density in dendritic branches of neurons \cite{HHIL}. This 
current density is directed parallel to the long axis of the dendrite and is 
driven by an electric field established by energy consuming cellular processes 
which vary in time as ion channels open and close in response to neuronal 
activity. The connection between the current density and the microscopic scale 
electric field which drives it is via Ohm's Law on the microscopic scale. For 
the sake of simplicity in this paper we will assume a static microscopic 
electric field, and therefore a steady current density, but generalization to 
a time varying field should be a relatively simple matter.

A model for the interaction of the acoustic field with the primary current will
presently be described. To present this model we will first work with the 
microscopic current and then, as indicated above, average it over $V_m$ to 
obtain the macroscopic primary current. The microscopic current density 
${\bf J}({\bf r})$ may be written as
\begin{equation}
{\bf J}({\bf r},t) = \rho^+({\bf r},t) {\bf v}^+({\bf r},t)
+ \rho^-({\bf r},t) {\bf v}^-({\bf r},t)
\label{pcaw1}
\end{equation}
where $\rho^\pm$ and ${\bf v}^\pm$ are the charge density and velocity 
respectively for the charge carriers. The velocity has a contribution from
charge drift ${\bf v}_d^\pm$ and bulk motion ${\bf v}_b$ so that ${\bf v}^\pm
= {\bf v}_d^\pm + {\bf v}_b$ and
\begin{equation}
{\bf J}({\bf r},t) = \rho^+({\bf r},t) {\bf v}_d^+({\bf r},t)
+ \rho^-({\bf r},t) {\bf v}_d^-({\bf r},t) 
+ [\rho^+({\bf r},t) + \rho^-({\bf r},t)] {\bf v}_b.
\label{pcaw1b}
\end{equation}
Using the relationship between drift velocity, mobility and electric field
we may write
\begin{equation}
{\bf J}({\bf r},t) = [\rho^+({\bf r},t) \mu^+({\bf r},t)
+ \rho^-({\bf r},t) \mu^-({\bf r},t)] {\bf E}({\bf r},t)
+ [\rho^+({\bf r},t) + \rho^-({\bf r},t)] {\bf v}_b.
\label{pcaw2}
\end{equation}
where ${\bf E}({\bf r},t)$ is the electric field on the microscopic scale and
$\mu^\pm$ is the mobility of the charge carriers. For the sake of simplicity
it has been assumed that there are only two ionic charge carriers but this
restriction could easily be relaxed. We also assume that the mobility is
isotropic and homogeneous. Relaxing this assumption (by assuming a tensor
mobility) could be a topic for further exploration. However since the current
is primarily directed along the long axis of dendrites in the brain then this
is not an unreasonable assumption with which to begin.

We will assume that prior to the application of the acoustic wave ${\bf v}_b=0$
but during the application of the wave ${\bf v}_b \ne 0$. Also when the 
acoustic wave is applied the charge density, charge mobility and electric 
fields are perturbed such that:
\begin{equation}
\rho^\pm = \rho^\pm_o + \Delta \rho^\pm
\qquad \mu^+ = \mu^+_o + \Delta \mu^+
\qquad {\bf E} = {\bf E}_o + \Delta {\bf E}
\label{pcaw3}
\end{equation}
where the first term on the right of each of these equations is the quantity
prior to application of acoustic wave and the second term is the small change 
due to the acoustic wave. 

Substituting Equations (\ref{pcaw3}) into Equation (\ref{pcaw2}) and keeping 
all but the small second order terms:
\begin{eqnarray}
{\bf J} &=& {\bf J}_o 
+ (\Delta \rho^+ \mu^+_o + \Delta \rho^- \mu^-_o) {\bf E}_o 
+ (\rho^+_o \Delta \mu^+ + \rho^-_o \Delta \mu^-) {\bf E}_o 
+ (\rho^+_o \mu^+_o + \rho^-_o \mu^-_o) \Delta {\bf E} 
\nonumber \\
&+& [\rho^+({\bf r},t) + \rho^-({\bf r},t)] {\bf v}_b
\label{pcaw4}
\end{eqnarray}
where the primary current prior to the application of the acoustic wave is
\begin{equation}
{\bf J}_o({\bf r},t) = (\rho^+_o \mu^+_o + \rho^-_o \mu^-_o) {\bf E}_o.
\label{pcaw5}
\end{equation}
According to Jossinet et al. \cite{Jossinet} for an adiabatic 
compression\footnote{The mechanical work done on a given volume of the fluid 
by the acoustic wave is converted into heat during a period of the wave. Since
the time constant associated with thermal diffusion is much larger than the 
period of the acoustic wave then the wave propagates without any thermal 
exchange with the local environment - i.e., adiabatically.} of a fluid we can
write
\begin{equation}
\Delta \rho^\pm  = \rho^+_o \beta_s \Delta p 
\label{pcaw5a}
\end{equation}
and
\begin{equation} 
\Delta \mu^\pm = \mu_o^\pm (H_p + m^\pm_T \Theta) \Delta p
\label{pcaw5b}
\end{equation}
where $\Delta p$ is the pressure change due to the acoustic wave and where
the thermodynamic constants $H_p$, $\beta_s$, $m^\pm_T$ and $\Theta$ are 
defined in \cite{Jossinet}. Therefore
\begin{eqnarray}
{\bf J} &=& {\bf J}_o 
+ \beta_s(\rho^+_o \mu^+_o + \rho^-_o \mu^-_o) {\bf E}_o \Delta p 
+ H_p(\rho^+_o \mu^+_o + \rho^-_o \mu^-_o) {\bf E}_o \Delta p
\nonumber \\
&+& (m^+_T \rho^+_o \mu^+_o
+ m^-_T \rho^-_o \mu^-_o) {\bf E}_o  \Theta \Delta p
+ (\rho^+_o \mu^+_o + \rho^-_o \mu^-_o) \Delta {\bf E} 
+ [\rho^+({\bf r},t) + \rho^-({\bf r},t)] {\bf v}_b
\label{pcaw6}
\end{eqnarray}
or
\begin{eqnarray}
{\bf J} 
&=& {\bf J}_o + (\beta_s + H_p) {\bf J}_o \Delta p
+ (m^+_T \rho^+_o \mu^+_o + m^-_T \rho^-_o \mu^-_o) {\bf E}_o \Theta \Delta p
\nonumber \\
&+& (\rho^+_o \mu^+_o + \rho^-_o \mu^-_o) \Delta {\bf E} 
+ [\rho^+({\bf r},t) + \rho^-({\bf r},t)] {\bf v}_b
\label{pcaw7}
\end{eqnarray}
In Equation (\ref{pcaw7}) the term that involves the change $\Delta {\bf E}$ 
in the microscopic electric field will now be investigated. The total 
microscopic charge density $\rho_t = \rho^+ + \rho^-$ (which establishes
the microscopic electric field driving the current along a dendrite) during 
application of the acoustic wave will be $\rho_t({\bf r}) = \rho_{to}({\bf r}) 
+ \Delta \rho_t({\bf r})$ where $\rho_{to}({\bf r})$ is the total charge 
density prior to application of the acoustic wave and $\Delta \rho_t({\bf r})$ 
is a small change in the total charge density due to the acoustic wave. Then
\begin{eqnarray}
 {\bf E}_o({\bf r}) = \int \rho_{to}({\bf r}')
 \frac{({\bf r} - {\bf r}')}{|{\bf r} - {\bf r}'|^3} \rmd^3 {\bf r}' 
 \qquad 
 \Delta {\bf E}({\bf r}) = \int \Delta \rho_t({\bf r}')
 \frac{({\bf r} - {\bf r}')}{|{\bf r} - {\bf r}'|^3} \rmd^3 {\bf r}'
\label{ef2a}
\end{eqnarray}
Since the total charge density is assumed to be zero on the macroscopic scale 
then to good approximation the electric field at any point ${\bf r}$ will be 
due to the microscopic distribution of charge arising locally within some small
volume $V_{\bf r}$ (possibly smaller than macroscopic volume $V_m$) about 
${\bf r}$ and Equation (\ref{ef2a}) becomes
\begin{eqnarray}
 {\bf E}_o({\bf r}) = \int_{V_{\bf r}} \rho_{to}({\bf r}')
 \frac{({\bf r} - {\bf r}')}{|{\bf r} - {\bf r}'|^3} \rmd^3 {\bf r}' 
 \qquad 
 \Delta {\bf E}({\bf r}) = \int_{V_{\bf r}} \Delta \rho_t({\bf r}')
 \frac{({\bf r} - {\bf r}')}{|{\bf r} - {\bf r}'|^3} \rmd^3 {\bf r}'
\label{ef2b}
\end{eqnarray}
Assume that the mass density of the acoustic medium  is $\rho({\bf r},0)$ 
when at rest and at equilibrium with external forces. When an acoustic 
wave disturbs this initially resting medium a small perturbation,  
$\Delta \rho({\bf r},t)$, to the mass density occurs so that 
\begin{equation}
\rho({\bf r},t) = \rho({\bf r},0) + \Delta \rho({\bf r},t).
\label{current_0}
\end{equation}
Since the drift velocity of the charge carriers can be safely assumed to be 
much less than the velocity of sound in the medium then the charge carriers 
will behave as though they are fixed relative to the medium on the time-scale 
of the acoustic wave propagation. In such a case the ratio of the mass density 
at time $t$ to that at time zero will be equal to the ratio of the charge 
density $\rho^{\pm}$ at time $t$ to that at time zero or:
\begin{equation}
\frac{\Delta \rho_t({\bf r})}{\rho_{to}({\bf r})} =
\frac{\Delta \rho({\bf r})}{\rho_o({\bf r})}
\label{ef3}
\end{equation}
then for uniform initial mass density $\rho_o({\bf r}) = \rho_o$
\begin{eqnarray}
 \Delta {\bf E}({\bf r}) = \frac{1}{\rho_o}
 \int_{V_{\bf r}} \rho_{to}({\bf r}') \Delta \rho({\bf r}') 
 \frac{({\bf r} - {\bf r}')}{|{\bf r} - {\bf r}'|^3} \rmd^3 {\bf r}'
 \label{ef5}
\end{eqnarray}
For a ${V_{\bf r}}$ on a scale less than the wavelength of the acoustic 
carrier frequency the mass density $\Delta \rho({\bf r}')$ will vary little 
over ${V_{\bf r}}$ and to good approximation the integral becomes
\begin{eqnarray}
 \Delta {\bf E}({\bf r}) &=& \frac{\Delta \rho({\bf r})}{\rho_o}
 \int_{V_{\bf r}} \rho_{to}({\bf r}') 
 \frac{({\bf r} - {\bf r}')}{|{\bf r} - {\bf r}'|^3} \rmd^3 {\bf r}'
 \label{ef5b} \\
 &=& \frac{\Delta \rho({\bf r})}{\rho_o} {\bf E}_o({\bf r})
 \label{ef6}
\end{eqnarray}

Combining Equations (\ref{pcaw7}) and (\ref{ef6}) 
using the relationship between $\Delta \rho$ and $\Delta p$ given
in Equation (\ref{pcaw5a}) we write
\begin{eqnarray}
 {\bf J} &=& {\bf J}_o 
 + (2 \beta_s + H_p) {\bf J}_o \Delta p
 + (m^+_T \rho^+_o \mu^+_o + m^-_T \rho^-_o \mu^-_o) {\bf E}_o  \Theta \Delta p
 + [\rho^+({\bf r},t) + \rho^-({\bf r},t)] {\bf v}_b
 \label{ef8}
\end{eqnarray}
To obtain the macroscopic current ${\overline {\bf J}}$ we average the 
microscopic current over a volume $V$ on the macroscopic spatial scale of 
interest to obtain
\begin{eqnarray}
 {\overline {\bf J}} &=& {\overline {\bf J}_o}
 + (2\beta_s + H_p) {\overline{{\bf J}_o \Delta p}}
 + \Theta (m^+_T \mu^+_o {\overline{\rho^+_o {\bf E}_o  \Delta p}}
 + m^-_T \mu^-_o  {\overline{\rho^-_o {\bf E}_o  \Delta p}})
 \\ \nonumber
 &+& {\overline{[\rho^+({\bf r},t) + \rho^-({\bf r},t)] {\bf v}_b}}
 \label{ef9}
\end{eqnarray}
where the horizontal line denotes an average over $V$. Since the bulk velocity
${\bf v}_b$ and acoustic pressure change $\Delta p$ vary little over $V$ (ie.
the acoustic wavelength is assumed to be much greater than the scale of $V$)
we can write
\begin{eqnarray}
 {\overline {\bf J}} &=& {\overline {\bf J}_o}
 + (2\beta_s + H_p) {\overline{{\bf J}_o}} \Delta p
 + \Theta (m^+_T \mu^+_o {\overline{\rho^+_o {\bf E}_o}}
 + m^-_T \mu^-_o  {\overline{\rho^-_o {\bf E}_o}})  \Delta p
 \\ \nonumber
 &+& {\overline{[\rho^+({\bf r},t) + \rho^-({\bf r},t)]}} {\bf v}_b
 \label{ef10}
\end{eqnarray}
and since the macroscopic charge density is assumed to remain zero at all 
times then
\begin{eqnarray}
 {\overline {\bf J}} &=& {\overline {\bf J}_o}
 + (2\beta_s + H_p) {\overline{{\bf J}_o}} \Delta p
 + \Theta (m^+_T \mu^+_o {\overline{\rho^+_o {\bf E}_o}}
 + m^-_T \mu^-_o  {\overline{\rho^-_o {\bf E}_o}})  \Delta p.
 \label{ef11}
\end{eqnarray}

Equation (\ref{ef11}) may also be conveniently written as
\begin{eqnarray}
 {\overline {\bf J}} &=& {\overline {\bf J}_o}
 + (2\beta_s + H_p + \Theta m_T) {\overline{{\bf J}_o}} \Delta p
 + \Theta (\delta_{m^+_T} {\overline{{\bf J}_o^+}}
 + \delta_{m^-_T} {\overline{{\bf J}_o^-}})  \Delta p.
 \label{ef11c}
\end{eqnarray}
where $m_T = (m^-_T + m^+_T)/2$, $\delta_{\mu^\pm} = m_T^\pm - m_T$ and
$\overline{{\bf J}_o^\pm}$ is the macroscopic current density due to the
signed charge carrier in the absence of the acoustic wave.
The last term on the right of Equation (\ref{ef11c}) is due to the unequal 
mobilities of the charge carriers. If the only charge carriers were sodium
and chloride ions then this term would be zero (see Table (\ref{table1})).
Regardless both contributions to the current density in the presence of the
applied acoustic wave will be due to neuronal activity. This motivates 
defining an {\it current density activity modulus} $\overline{{\bf J}_a}$ 
according to
\begin{eqnarray}
 {\overline {\bf J}_a} &=& (2\beta_s + H_p) {\overline{{\bf J}_o}} 
 + \Theta (m^+_T \mu^+_o {\overline{\rho^+_o {\bf E}_o}} 
 + m^-_T \mu^-_o  {\overline{\rho^-_o {\bf E}_o}}) 
 \label{ef12}
\end{eqnarray}
We then have
\begin{eqnarray}
 {\overline {\bf J}} 
 &=& {\overline {\bf J}_o} + {\overline{{\bf J}_a}} \Delta p.
 \label{ef13}
\end{eqnarray}

We will assume that the medium is liquid-like and therefore only supports 
longitudinal acoustic waves. Such a spherical mass density wave propagating 
through an acoustic medium with velocity $v$ may be written as\footnote{In 
human soft tissue the velocity of sound varies little with respect to tissue 
type and is usually assumed to be that of water, 1,482 m/s at $20^{\circ}$ C, 
in medical diagnostic ultrasound applications.}
\begin{equation}
\Delta p({\bf r},t) = \frac{1}{r} \psi(r \pm v t).
\label{ef14}
\end{equation}
Combining Equations (\ref{ef13}) and (\ref{ef14}) one obtains
\begin{equation}
\overline{\bf J}({\bf r},t) =  \overline{\bf J}_o({\bf r}) 
+  \frac{1}{r} \psi(r \pm v t) \overline{\bf J}_a(\bf r)
\label{ef15}
\end{equation}
as the current density in the presence of the applied spherical acoustic wave.
In the next section a relationship between the coefficients of a vector 
spherical harmonic expansion of the current density activity modulus 
$\overline{{\bf J}_a}$ and the measured magnetic field will be derived. This
relationship will be the imaging equation which when inverted yields an image 
of the non-silent part of the current density activity modulus. 

\begin{table}[ht]
 \centering
 \renewcommand{\arraystretch}{1.4}
 \begin{tabular}{|c|c|}
  \hline
  $\beta_s$ & 4.56 x $10^{-10}$ Pa\textsuperscript{-1} \\
  \hline
  $H_p$ & -1.73 x $10^{-10}$ Pa \textsuperscript{-1} \\ 
  \hline
  $\Theta$ & 1.4 x $10^{-8}$ K Pa\textsuperscript{-1} \\
  \hline
  $m_T$ (Na\textsuperscript{+}) & 2.48 $10^{-2}$ K\textsuperscript{-1} \\
  \hline
  $m_T$ (K\textsuperscript{+}) & 2.17 $10^{-2}$ K\textsuperscript{-1} \\
  \hline
  $m_T$ (Cl\textsuperscript{-}) & 2.48 $10^{-2}$ K\textsuperscript{-1} \\
  \hline
 \end{tabular}
 \caption{Thermodynamic constants \cite{Jossinet}}.
 \label{table1}
\end{table}

\section{The Magnetic Field Data}
\label{mag_field}

In this section we develop a near-field (quasistatic) equation which relates 
the measured magnetic field to the current density activity modulus. Since 
Maxwell's Laws are linear with respect to the current source then the first 
and second terms on the right side of Equation (\ref{ef15}) will respectively 
result in temporally static and dynamic contributions to the measured magnetic 
field. The static contribution to the measured field can be removed by 
filtering or subtraction methods (see Section \ref{nonsilent}). We may 
therefore omit the contribution of the static current term to the measured 
field and consider only the {\it effective current} given by:
\begin{equation}
 {\bf J}_e({\bf r},t)
 = \frac{1}{r} \psi(r \pm v t) {\overline{\bf J}}_a(\bf r).
\label{current_13}
\end{equation}

We will assume that measurements of the magnetic field are made on a 
spherical shell of radius $R$ enclosing the compactly supported current 
density. It is then natural to expand the magnetic field (and vector potential)
in terms of vector spherical harmonics (VSH). The resulting expressions will 
allow a precise characterization of the quantities associated with the current 
density which can be determined uniquely. External to the current containing 
region we may write \cite{JDJ}:       
\begin{equation}
{\bf A}({\bf r},t) =
\frac{1}{c} \int \int
\frac{{\bf J}_e({\bf r}',t')}{|{\bf r}' - {\bf r}|}  
\delta \left(t' + \frac{|{\bf r} - {\bf r}'|}{c} - t \right)
\rmd^{3}{\bf r}^{\prime} dt'.
\label{near_field_1}
\end{equation}
where $c$ is the speed of light. Using the current given by Equation 
(\ref{current_13}) this becomes
\begin{equation}
 {\bf A}({\bf r},t) = \frac{1}{c}
 \int 
 \left[ 
   \int \frac{1}{r'} \psi(r' \pm v t') 
   \delta \left(t' + \frac{|{\bf r} - {\bf r}'|}{c} - t \right) dt' 
 \right]
 \frac{{\overline{\bf J}}_a({\bf r}')}{|{\bf r}' - {\bf r}|} 
 \rmd^{3}{\bf r}^{\prime} 
 \label{near_field_2}
\end{equation}
and performing the temporal integration leads to
\begin{equation}
 {\bf A}({\bf r},t) = \frac{1}{c}
 \int \frac{1}{r'} 
 \psi(r' \pm vt - \beta|{\bf r} - {\bf r}'|) 
 \frac{{\overline{\bf J}}_a({\bf r}')}{|{\bf r}' - {\bf r}|} 
 \rmd^{3}{\bf r}^{\prime} 
 \label{near_field_3a}
\end{equation}
where $\beta = v/c$. The speed of sound (water = 1482 m/s, animal tissue = 
1540 m/s) is much less than the speed of light (3.0 x $10^{8}$ m/s) so that 
$\beta \ll 1$ and since $|{\bf r} - {\bf r}'| < R$ we can to excellent
approximation write
\begin{equation}
 {\bf A}({\bf r},t) = \frac{1}{c}
 \int \frac{1}{r'} \psi(r' \pm vt) 
 \frac{{\overline{\bf J}}_a({\bf r}')}{|{\bf r}' - {\bf r}|} 
 \rmd^{3}{\bf r}^{\prime} 
 \label{near_field_3e}
\end{equation}
which is the near-field (or quasistatic) approximation.

Using a vector spherical harmonic (VSH) expansion \cite{VMK} of the infinite 
space Green's function for the Laplace equation (see appendix \ref{sec-sphere})
and Equation (\ref{near_field_3e}) we may write the field outside of the sphere
supporting the current density as
\begin{equation}
 {\bf A}({\bf r},t) = \frac{4 \pi}{c} \sum_{ljm} 
 \left( 
   \int \frac{{r'}^{l-1}}{2l+1} \psi(r' \pm vt) {\overline{\bf J}}_a({\bf r}') 
   \cdot {\bf Y}_{jm}^{*l}(\Omega') \rmd^{3}{\bf r}^{\prime} 
 \right)
 \frac{{\bf Y}_{jm}^{l}(\Omega)}{r^{l+1}}.
 \label{megns1}
\end{equation}
where $j=0, \ldots, \infty $, $ l=j,j+1,j-1 $ (with the exception that 
$l=1$ for $j=0$), and $m=-j,-j+1,...,j-1,j$. We will, as in the above equation,
use the variable $\Omega$ to denote the ordered pair of angular
variables $(\theta, \phi)$. 

If we write ${\overline{\bf J}}_a({\bf r})$ in a VSH expansion (see Appendix
\ref{sec-sphere}) as
\begin{equation}
 {\overline{\bf J}}_a({\bf r'}) =  
 \sum_{l'j'm'} \alpha^{l'}_{j'm'}(r') {\bf Y}^{l'}_{j'm'}(\Omega')
 \label{megns1b}
\end{equation}
and substitute into Equation (\ref{megns1}) then we arrive at
\begin{equation}
 {\bf A}({\bf r},t) =  \frac{4 \pi}{c} \sum_{ljm}
 \left( 
   \frac{\int_0^R r'^{l+1} \psi(r' \pm vt) \alpha_{jm}^{l}(r') \rmd r'}{2l+1} 
 \right)
 \frac{{\bf Y}_{jm}^{l}(\Omega)}{{r}^{l+1}}
 \label{megns2}
\end{equation}

Taking the curl of Equation (\ref{megns2}) (see Appendix \ref{sec-sphere}) and 
using the condition that $\nabla \times {\bf B} = 0$ outside the region of 
support for the current density then the magnetic field on the spherical 
measurement shell is:
\begin{eqnarray}
 {\bf B}(\Omega, t) 
 &=& -\rmi \frac{4 \pi}{c} \sum_{jm} 
 \left( \frac{j}{2j+1} \right)^{1/2} 
 R^{-(j+2)} {\bf Y}_{jm}^{j+1}(\Omega)
 \int_{0}^{R} r'^{j+1} 
 \psi(r' \pm vt) \alpha_{jm}^{j}(r') \rmd r'. 
 \label{megns5}
\end{eqnarray}   
Therefore all ${\bf Y}_{jm}^{j+1}$ and  ${\bf Y}_{jm}^{j-1}$ components of the 
current are silent in that they produce no field on the measurement shell. 
In addition any ${\bf Y}_{jm}^j$ component of the current for which the radial
integral in Equation (\ref{megns5}) is zero will also be silent.

It is well known that for regions where the current density vanishes a scalar 
magnetic field potential $\Phi$ exists such that ${\bf B}=-\nabla \Phi$ and 
which obeys the Laplace equation $\nabla^2\Phi = 0$. Therefore, according to
standard potential theory, a solution of the Laplace equation outside of the
sphere can be obtained from knowledge of the normal derivative of the scalar
potential on the surface enclosing the volume. Therefore all that is needed 
to fully determine ${\bf B}$ outside the measurement sphere is the knowledge 
of $B_r$ on the surface of the sphere. Using the definitions of the VSH given 
in appendix \ref{sec-sphere} one easily obtains $B_r$:
\begin{equation}
 B_{r}(\Omega, t) = \rmi \frac{4 \pi}{c} \sum_{jm}
 \frac{\sqrt{j(j+1)}}{2j+1} R^{-(j+2)}
 \int_{0}^{R} r'^{j+1} \psi(r' \pm vt) \alpha_{jm}^{j}(r') \rmd r' \:
 Y_{jm}(\Omega)
 \label{megns6}
\end{equation} 
and we arrive at
\begin{equation}
 {\mathcal B}_{jm}(t)
 = \int_{0}^{R} r'^{j+1} \psi(r' \pm vt) \alpha_{jm}^{j}(r') \rmd r' 
 \label{megns7}
\end{equation}
where we have defined ${\mathcal B}_{jm}(t) = -\rmi \frac{c}{4\pi}
\frac{2j+1}{\sqrt{j(j+1)}} R^{j+2} B_{jm}(t)$ and $B_{jm}(t) = \int B_{r}
(\Omega, t) Y_{jm}^*(\Omega) \rmd \Omega $. This equation, the basic result of 
this section, connects the 
external magnetic field measurement in the presence of a spatially selective 
spherical acoustic wave to a single radial moment of each coefficient 
$\alpha_{jm}^{j}(r)$. 

Before moving on to the inversion of the magnetic field data it may be
instructive to note that Equation (\ref{megns7}) gives insight into the extent 
to which the inverse problem of MEG ($\psi = 1$) is ill-posed. An 
$N$-{\it point radial digital resolution} will be defined by the the value of 
$\alpha_{jm}^j(r)$, and hence the current density image, at $N-1$ equidistant 
points $r_n$ in the radial direction. For example, a $10$-point radial 
resolution within a measurement shell of radius 10 cm would correspond to a 
1.0 cm radial resolution.  Let us assume that the current density is comprised 
of ${\bf Y}_{jm}^j$ components only (i.e., there are no silent parts of the 
current density) and that all $\alpha_{jm}^j(r)$ are well approximated by a 
$N^{th}$ order polynomial. To determine such a polynomial from its moments
would require the knowledge of $N+1$ moments. From Equation (\ref{megns7}) we 
see that the field measurements give one moment only for each 
$\alpha_{jm}^j(r)$. One can then state that for a unique $N$-point resolution 
inversion $1/(N+1)$ of the data is obtained from the field measurements and 
$N/N+1$ of the data is obtained from the additional constraints. Then, 
practically speaking, to obtain a 1.0 cm radial digital resolution within a 
sphere of radius $R=10$ cm only 10 percent of the data necessary for a unique 
inversion would come from the field measurement alone!

\section{Inverting for the Non-Silent Current Density}
\label{nonsilent}

In this section we investigate the use of Equation (\ref{megns7}) in 
determining the coefficients $\alpha_{jm}^j(r)$ and therefore the current 
density activity modulus. For the remainder of this paper the outgoing
wave (ie. argument is $r-vt$) will be assumed (An ingoing wave would
be treated similarly). Since any physical acoustic wave must be causal then 
we must have $\psi(r'-vt) = 0$ for $r'>vt$ so that Equation (\ref{megns7}) 
becomes
\begin{eqnarray}
 {\mathcal B}_{jm}(t) = 
 \int_{0}^{vt} \psi(r'-vt) r'^{j+1} \alpha_{jm}^{j}(r') \rmd r' 
 \qquad \qquad 0<t<\frac{R}{v}.
 \label{unique_100a}
\end{eqnarray}
Setting $r=vt$ (The acoustic wave speed $v$ is assumed to be known.) and 
defining the continuously measured data ${\mathcal D}_{jm}(r) = 
{\mathcal B}_{jm}(r/v)$ then Equation (\ref{unique_100a}) can be written as 
\begin{eqnarray}
 {\mathcal D}_{jm}(r) = 
 \int_0^r \psi(r'-r) g_{jm}(r') \rmd r'
 \qquad \qquad 0<r<R
 \label{unique_100b}
\end{eqnarray}
where $g_{jm}(r') = r'^{j+1} \alpha_{jm}^{j}(r')$.

Equation (\ref{unique_100b}) is a Volterra integral equation of the first 
kind \cite{Poly} with convolution-type kernel. It relates the measured data 
${\mathcal D}_{jm}(r)$ to the unknown function $g_{jm}(r')$ on the interval 
$0 \le r' \le R$. Well established methods of solving for $g_{jm}$ can be 
found in the literature (see for example \cite{Hansen,Lamm,Poly,Wing}). The 
solutions are known to be unique if: (1) $D_{jm}(r')$ is continuously 
differentiable on $0 \le r' \le R$ and $D_{jm}(0)=0$, (2) $\Psi(r'-r)$ and 
$\partial \Psi(r'-r)/\partial r$ are continuous on $0 \le r' \le r \le R$, and 
(3) $\psi(0) \ne 0$. 

Although the solution of Equation (\ref{unique_100b}) is unique under expected 
conditions the solution is still ill-posed in the sense that it can be very 
sensitive to noise in the data ${\mathcal D}_{jm}(r)$. Even though the 
underlying signal is expected to be differentiable on $0 \le r' \le R$ the 
addition of a noise component to the measured signal will create a 
${\mathcal D}_{jm}(r)$ that in general will not be differentiable. The 
violation of condition (1) is known to lead to an amplification of high 
frequency noise and the degree to which the condition is violated controls the 
amplification. Signal averaging and smoothing of the data can be used to 
diminish the effect of noise amplification but usually some type of 
regularization (Tikhonov regularization is common.) will be needed. Furthermore
the usual consequence of regularization is that resolution decreases as 
regularization is increased.

To solve Equation (\ref{unique_100b}) numerical methods are employed which take
as input the data ${\mathcal D}_{jm}(r_n)$ sampled at discrete points $r_n = 
n \Delta r = nv \Delta t$ where $\Delta t$ is the intersample interval. It can 
be shown \cite{Hack} that solutions obtained through the discretization of the 
integral equation tend toward the exact solution as $\Delta t \rightarrow 0$. 
Therefore one might then think it advantageous to choose $\Delta t$ a small as 
possible. However a decrease in $\Delta t$ in the presence of data noise is 
expected to exacerbate the violation of condition (1) and increase sensitivity
to noise. The above discussion highlights the difficulty in obtaining a
theoretical bound on the radial resolution of the proposed method but doesn't
prevent establishing resolution by empirical means. 

It may be advantageous to modulate the acoustic wave with a carrier frequency 
$k_o$ so that
\begin{equation}
 \psi(r - vt) = \Omega(r - vt) e^{i2\pi(r - vt)k_o}.
 \label{unique_103}
\end{equation}
This will have the effect of shifting the frequency spectrum of the signal and 
allow for the filtering of the DC component given by ${\overline {\bf J}_o}$ 
as well as low frequency noise. The choice of the wave envelope 
$\Omega(r - vt)$ will undoubtedly have consequences with respect to the
numerical inversion of Equation (\ref{unique_100b}) but this will only
briefly be addressed along with signal and noise considerations in the next
section.

\section{Signal, Noise and Resolution}  
\label{sec-sigmag}

In this section the signal and noise of the proposed method will 
be considered in rather broad terms. It is useful to estimate the signal 
strength of the proposed method (after filtering out the contribution of the 
DC term) compared to that of MEG. The ratio $R_{sig}$ of the two signals would 
be given by  
\begin{eqnarray}
 R_{sig} = \frac{|{\overline {\bf J}_a} \Delta p|}{|\overline{{\bf J}_o}|}.
 \label{snr1}
\end{eqnarray}
From Table \ref{table1} note that the difference between $m_T^+$ and $m_T^-$ 
for the most prevalent charge carriers in biological systems is about 10 
percent. It is therefore justified to use the following approximation to the 
current density activity modulus to estimate $R_{sig}$:
\begin{eqnarray}
 {\overline {\bf J}_a} 
 \approx (2\beta_s + H_p + \Theta m_T) {\overline{{\bf J}_o}}.
 \label{snr2}
\end{eqnarray}
(Note that Equation (\ref{snr2}) would be exact if sodium and chloride ions are 
the only charge carriers.). Then
\begin{eqnarray}
 R_{sig} \approx |(2\beta_s + H_p + \Theta m_T) \Delta p|.
 \label{snr3}
\end{eqnarray}

In diagnostic medical ultrasound imaging the overpressure $\Delta p$ is 
typically in the range 0.5-5.0 MPa. Using the values given Table \ref{table1} 
an estimate of $R_{sig} \approx$ 5.0 x $10^{-3}$ is obtained for $\Delta p =$
5.0 MPa. Therefore the signal from any spherical shell of RMS thickness 
$\sigma$ will be between 2 and 3 orders of magnitude smaller than that from 
the same shell in an MEG measurement. Since the signal strength of a 
neuromagnetic signal in MEG is typically in the range of 50-500 fT this is 
likely to present a challenge with respect to magnetometer sensitivities even 
with signal averaging. Whether larger overpressures could be used safely 
(avoiding cavitation), perhaps at higher carrier frequencies, to boost the 
signal strength without causing neuronal stimulation \cite{Hameroff} could be 
an area of experimental exploration. Additionally attenuation of the acoustic 
wave would be expected to increase with frequency presenting yet another 
challenge to obtaining a signal of sufficient strength. Note that in a model 
system built to demonstrate the proposed method perhaps $R_{sig}$ could be
made much larger than that estimated above.

Sources of magnetic and electronic noise in the proposed method may originate 
in the environment, the magnetometer, the magnetometer electronics and the 
body. For example environmental sources may include the acoustic wave source, 
communications systems (radio, television, and microwave transmitters), 
electric pumps, electric motors, large nearby moving objects (automobiles and 
elevators), and power distribution systems. Magnetometer related noise may be 
due to thermal noise in the magnetometer (for example, SQUID noise), noise in 
the supporting electronics and thermal Dewar noise. The electrical activity of 
neighboring organs such as muscle also generate a time varying magnetic field 
that may be considered to be noise. Even background brain alpha rhythms might 
be considered to be "noise" if averaging (to increase signal-to-noise ratio) 
is used in an experiment which is time-locked to a stimulus. There are many 
means by which these sources of noise may be reduced and once all means of 
reducing environmental noise are exhausted the ultimate remaining noise source 
will be that due to the thermally induced magnetic noise in the brain and body 
tissues. This noise has been estimated to be on the order of 0.1 
$\rm{fT}/\sqrt{\rm{Hz}}$ \cite{Varpula}.

If no carrier wave is used then the $\overline {\bf J}_o$ term in Equation 
(\ref{ef13}) might be removed by a subtraction method in which the signal is 
acquired in the presence and absence of the applied acoustic wave and the 
signals are subsequently subtracted to obtain the desired $\overline {\bf J}_a$
term of Equation (\ref{ef13}). If a carrier frequency is used then the 
$\overline {\bf J}_o$ term of Equation (\ref{ef13}) could be removed by 
high-pass filtering the temporal signal prior to deconvolution. Using a 
carrier frequency could potentially have another benefit in that the signal 
could be pushed into a frequency band significantly different from 
environmental noise sources and then be band-pass filtered to obtain a signal 
of reduced noise compared to that of the typical MEG experiment. However, 
the benefit of a carrier frequency may come at a cost. The assumption made in 
going from Equation (\ref{ef5}) to Equation (\ref{ef5b}) involves the wave 
length of the acoustic wave and will be increasingly difficult to meet at 
smaller wavelengths associated with a carrier wave.

\section{Discussion}  
\label{sec-disc}

We presently lack a relatively noninvasive tool to study intact brain activity
on the spatial and temporal scales associated with most brain function.
This paper explores one means by which a finer scale information of brain 
function might someday be accessible. The spirit of this work is to further the 
discussion of novel methods for relatively noninvasive studies of human brain 
activity. The author acknowledges that there would be many hurdles to achieving
this goal by the proposed method, even for the simple geometry and assumptions
considered here, and this paper does not present a comprehensive list of all 
such challenges. Instead this paper proposes a simple brain-like model for the 
interaction of an acoustic wave with a current carrying medium and shows how 
that interaction might potentially achieve the mapping of the non-silent
component of the current density on the scales of interest to neuroscience. 

In addition this work explores the interplay between signal, noise, 
safety, and inversion method that would ultimately place fundamental 
limitations on the proposed method. Such interplay is one known to all methods 
of imaging in the biological sciences. It is the hope of the author that this 
work and the work of those mentioned in this paper's introduction will 
stimulate others to think about methods by which current density could be 
spatially encoded on time and spatial scales needed to fruitfully and 
safely probe brain function.

\appendix

\section{Vector Spherical Harmonics}
\label{sec-sphere}
 
For convenience this appendix gives the definition and some properties of 
vector spherical harmonics \cite{VMK}. The vector spherical harmonics may be 
defined and generated from the scalar spherical harmonics according to: 
\begin{eqnarray}
{\bf Y}_{jm}^{j+1} & = &
\sqrt{\frac{j+1}{2j+1}}
\left( -{\bf e}_r Y_{jm} + {\bf e}_\theta \frac{1}{j+1}
\frac{\partial Y_{jm}}{\partial \theta} + {\bf e}_\phi \frac{im}{j+1} \frac{Y_{jm}}{\sin \theta} \right)
\nonumber \\
{\bf Y}_{jm}^{j} & = &
- {\bf e}_{\theta} \frac{m}{\sqrt{j(j+1)}} \frac{Y_{jm}}{\sin \theta}
- {\bf e}_{\phi} \frac{i}{\sqrt{j(j+1)}}
\frac{\partial Y_{jm}}{\partial \theta}
\nonumber \\
{\bf Y}_{jm}^{j-1} & = &
\sqrt{\frac{j}{2j+1}}
\left( {\bf e}_r  Y_{jm}
+ {\bf e}_{\theta} \frac{1}{j}
\frac{\partial Y_{jm}}{\partial \theta} \right.
\nonumber + \left. {\bf e}_{\phi} \frac{im}{j}
\frac{Y_{jm}}{\sin \theta} \right).
\nonumber
\end{eqnarray}

\noindent
The vector spherical harmonics obey the orthogonality property
 
\begin{equation}
\int_0^{\pi} \int_0^{2\pi}
{\bf Y}_{j^\prime m^\prime}^{*l^\prime} 
{\bf Y}_{jm}^l \sin \theta d\theta d\phi
= \delta_{jj^\prime} \delta_{ll^\prime} \delta_{mm^\prime}
\label{Vortho}
\end{equation}
 
\noindent
and the following relations for the divergence operator:    

\begin{eqnarray}
\nabla \cdot \left[ f(r) {\bf Y}_{jm}^{j+1} \right] &=&
- \sqrt{\frac{j+1}{2j+1}} \left( \frac{d}{dr} + \frac{j+2}{r} 
\right) f(r) Y_{jm} \qquad\\
\nabla \cdot \left[ f(r) {\bf Y}_{jm}^j  \right] &=& 0
\nonumber \\
\nabla \cdot \left[ f(r) {\bf Y}_{jm}^{j-1} \right] &=&
\sqrt{\frac{j}{2j+1}} \left( \frac{d}{dr} - \frac{j-1}{r} \right)
f(r) Y_{jm} \nonumber
\end{eqnarray}
 
\noindent and the curl operator:
 
\begin{eqnarray}
\label{Vcurl}
\nabla \times \left[ f(r) {\bf Y}_{jm}^{j+1} \right] &=&
i \sqrt{\frac{j}{2j+1}} \left( \frac{d}{dr} + \frac{j+2}{r} \right)
f(r) {\bf Y}_{jm}^j \qquad \: \\
\nabla \times \left[ f(r) {\bf Y}_{jm}^j \right] &=&
i \sqrt{\frac{j}{2j+1}} \left( \frac{d}{dr} - \frac{j}{r} \right)
f(r) {\bf Y}_{jm}^{j+1}
\nonumber \\
&+& i \sqrt{\frac{j+1}{2j+1}} \left( \frac{d}{dr} +
\frac{j+1}{r} \right)
f(r) {\bf Y}_{jm}^{j-1}
\nonumber \\
\nabla \times \left[ f(r) {\bf Y}_{jm}^{j-1} \right] &=&
i \sqrt{\frac{j+1}{2j+1}} \left( \frac{d}{dr} - \frac{j-1}{r} \right)
f(r) {\bf Y}_{jm}^j.
\nonumber
\end{eqnarray}            

Since the vector spherical harmonics are complete on the space of functions for which $\int |\mathbf{F}(\Omega)|^2 d\Omega < \infty$ (i.e., square integrable 
functions of $\Omega$) then any function this space can be expanded as:
\begin{eqnarray}
\mathbf{F}(\mathbf{r})
= \sum_{ljm} \alpha_{jm}^l(r) {\bf Y}_{jm}^{l}(\Omega).
\nonumber
\end{eqnarray}
In particular the Green's function for the infinite space Laplace equation can 
be written as
\begin{equation}
\frac{{\bf J}({\bf r'})}{|{\bf r}' - {\bf r}|} = 4\pi
\sum_{ljm} \left(
\frac{{r'}^l}{2l+1} {\bf J}({\bf r}') \cdot
{\bf Y}_{jm}^{*l}(\Omega') \rmd^{3}{\bf r}^{\prime} \right)
\frac{{\bf Y}_{jm}^{l}(\Omega)}{r^{(l+1)}}
\nonumber
\end{equation}
where it has been assumed that $r > r'$.

\bibliography{mag_res}

\end{document}